\documentclass[12pt]{iopart}
\usepackage{iopams}
\usepackage{appendix}
\usepackage{graphicx}
\usepackage{subfigure}
\usepackage{dcolumn}
\usepackage{bm}
\usepackage{epstopdf}
\usepackage{xcolor}
\usepackage{footmisc}
\usepackage{hyperref}
\expandafter\let\csname equation*\endcsname\relax
\expandafter\let\csname endequation*\endcsname\relax
\usepackage{amsmath}

\begin{document}

\title[QCQ layout Design Optimization and Automation]{Quantum Chip Design Optimization and Automation in Superconducting Coupler Architecture}

\author{Fei-Yu Li and Li-Jing Jin}

\address{Institute for Quantum Computing, Baidu Research, Beijing, China}
\ead{\textcolor{blue}{jinlijing2049@outlook.com}}
\vspace{10pt}

\begin{abstract}
Superconducting coupler architecture demonstrates great potential for scalable and high-performance quantum processors, yet how to design efficiently and automatically ``Qubit-Coupler-Qubit (QCQ)" of high performance from the layout perspective remains obscure.
In this work, this issue is studied for the first time resulting in three key findings.
Firstly, we acquire the crucial
zero-coupling condition that is only dependent on the geometric design of the layout. 
Secondly, the upper bound of the qubit-qubit effective coupling is found as $0.0822~ \omega_l/\beta_s^2$ which surprisingly depends only on the artificially pre-decided quantities $\omega_l, \beta_s$ instead of specific layouts. 
Thirdly, we propose an optimal layout design procedure to reach the very upper bound, leading to efficient and high-performance layout design. The effectiveness of the procedure has been demonstrated scrupulously using electromagnetic simulation experiments. As a stirring application, we report a state-of-the-art 3202 um long-range and scalable QCQ layout that is especially crucial to quantum error correction.
Our work provides practical guides to optimize the performance of the existing coupler architecture, find out novel layouts, and further advance the progress of quantum chip design automation.
\end{abstract}

\noindent{\it Keywords: \/}superconducting quantum chip, tunable coupler, layout design, optimization, automation
%
%
%
%
%

\section{Introduction}

Since the last century, several remarkable quantum algorithms such as Shor's algorithm \cite{shorAlgorithmsQuantumComputation1994} and Grover's algorithm \cite{groverFastQuantumMechanical1996a} have been proposed to solve certain problems which are essentially inaccessible for current classical computers, revealing the overwhelming capability of quantum computation. As one of the promising physical realizations for quantum computation, superconducting circuits draw people's attention due to their better manipulability and scalability and have been developed intensively in the past two decades, see comprehensive reviews in \cite{krantzQuantumEngineerGuide2019,clarkeSuperconductingQuantumBits2008,youSuperconductingCircuitsQuantum2005,youAtomicPhysicsQuantum2011,oliverMaterialsSuperconductingQuantum2013,devoretImplementingQubitsSuperconducting2004,guMicrowavePhotonicsSuperconducting2017,girvinCircuitQEDSuperconducting2014,schoelkopfWiringQuantumSystems2008,gambettaBuildingLogicalQubits2017,wendinQuantumInformationProcessing2017,gaoPracticalGuideBuilding2021}. As the kernel of superconducting quantum computation, the quantum processor through which quantum gates are implemented to realize quantum algorithms plays a critical role. In order to realize high-fidelity two-qubit gates which are
crucial and challenging for scalable quantum computing, Yan et al. \cite{yanTunableCouplingScheme2018} proposed a novel tunable coupling architecture that can be used to completely turn off the coupling (i.e., ``zero-coupling") between neighboring qubits realizing a strong suppression of parasitic effects. Thanks to this architecture, the performance of superconducting quantum processors reaches a brand new level. As an impressive application, quantum advantages have been successfully demonstrated experimentally \cite{aruteQuantumSupremacyUsing2019,wuStrongQuantumComputational2021,zhuQuantumComputationalAdvantage2022}. Despite the enlightenment of this architecture, however, challenges remain when it comes to the practical layout design. First and foremost, stronger qubit-qubit effective (QQe) couplings are preferred for high-fidelity two-qubit gates \cite{ashhabSpeedLimitsQuantum2012} when the coupling is turned on at a certain coupler frequency $\omega_{\text{on}}$, yet obstructed by various limitations. A primary one is the dispersive coupling between qubits and tunable couplers, which usually acts as a precondition in coupler architecture. Also, it is nontrivial to attain an optimal layout that can realize a strong QQe coupling at $\omega_{\text{on}}$ and reach zero-coupling at the other coupler frequency $\omega_{\text{off}}$ simultaneously. Together with other practical limitations (e.g., Micro and nanotechnology), a delicate and efficient layout design procedure instead of laborious trials to enable a quantum processor of high performance is desired since no well-defined procedure exists yet to our knowledge, which significantly retards the development of quantum chip design optimization and automation. 
In particular, as an attractive and prospective application, the long-range QCQ layout design \cite{seteFloatingTunableCoupler2021,marxerLongdistanceTransmonCoupler2022} is notably faced with these challenges.

To address these challenges, we concentrate on the layout of the QCQ architecture specifically and build an inspired layout design procedure for stronger QQe coupling under certain restrictions.
Normally, the QCQ layout contains two main components: the geometric metal layout (m-layout) including element polar plates as well as the surrounding ground, and the Josephson junction layout (JJ-layout).
Using circuit Quantum Electrodynamics (cQED) theory, we obtain three key findings that are significantly useful to the QCQ layout design.
Firstly, we discover the general zero-coupling condition for neighboring qubits. To our surprise, it depends only on the geometric design of the m-layout. Secondly, a quantitative upper bound of QQe couplings is found. It is only restricted by the qubit-coupler dispersive coupling and limited element frequency, but independent of the specific layout design. Last but not least, to achieve the maximum QQe coupling, we propose an optimal procedure for QCQ layout design. Making use of this procedure, we offer practical QCQ layouts as examples and demonstrate their performance using electromagnetic simulation experiments. As a stirring and promising application, a 3202-um long-range and scalable QCQ layout is designed successfully with aid of these findings.

\section{The model}

\begin{figure*}[htbp]
\centering
\includegraphics[width=\linewidth]{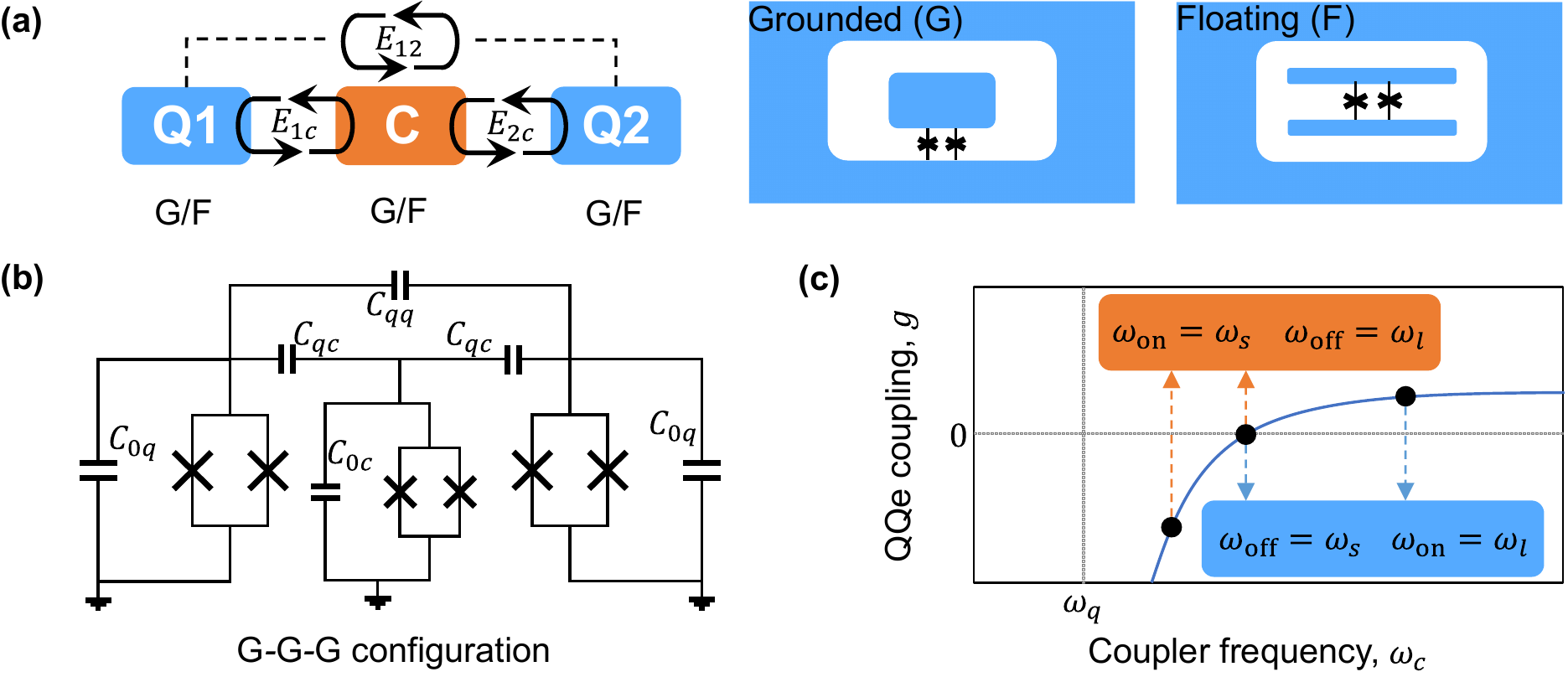}
\caption{ (Color online) (a) (left) The QCQ configuration. (right) The layout diagrams of grounded (G) type and floating (F) types, where the region in blue represents superconducting metal and SQUIDs are connected between metal plates. (b) The equivalent circuit diagram of the QCQ system in G-G-G configuration, described in terms of $\mathcal{C}=\left\{C_{0q}, C_{0c}, C_{qc}, C_{qq}\right\}$. (c) QQe coupling characteristics in two different sub-regimes: $\omega_{\text{on}} < \omega_{\text{off}}$ (orange) and $\omega_{\text{on}} > \omega_{\text{off}}$ (blue), where $\omega_{\text{off}}$ represents the zero-coupling between qubits and $\omega_{\text{on}}$ corresponds to a switching on point.}
\label{circuits}
\end{figure*}

The QCQ architecture based on capacitive couplings has several different configurations due to different structures of the qubit and coupler \cite{yanTunableCouplingScheme2018,seteFloatingTunableCoupler2021,aruteQuantumSupremacyUsing2019,zhuQuantumComputationalAdvantage2022,wuStrongQuantumComputational2021,zhangManybodyHilbertSpace2022a,xuHighFidelityHighScalabilityTwoQubit2020,collodoImplementationConditionalPhase2020,liTunableCouplerRealizing2020,sungRealizationHighFidelityCZ2021,stehlikTunableCouplingArchitecture2021} as shown in Figure~\ref{circuits}(a). Two main layout structures of qubit and coupler are distinguished as grounded (G) type and floating (F) type. Nevertheless, all these QCQ configurations share a uniform quantized Hamiltonian (see \ref{section1})
\begin{equation} \label{ham0}
        \hat{H} =  \sum_{k\in \{1,2,c\}} 4E_{Ck}\hat{n}_k^2 -E_{Jk}\cos(\hat{\phi}_k) 
        + 4E_{1c}\hat{n}_1\hat{n}_c + 4E_{2c}\hat{n}_2\hat{n}_c + 4E_{12}\hat{n}_1\hat{n}_2, 
\end{equation}
where the indices $1, 2, c$ are denoted as two qubits and coupler respectively; $\hat{n}_k=\hat{Q}_k/2e$, $\hat{\phi}_k=2\pi\hat{\Phi}_k/\Phi_0$, $E_{Ck}$, $E_{Jk}$ ($k\in\{1,2,c\}$) are Cooper-pair number operators, reduced flux operators (or phase operators), charging energies and Josephson energies respectively with $Q_k, \Phi_k, \Phi_0=h/2e$ are charges on the capacitances, fluxes through the Josephson junctions and the flux quantum respectively; $E_{1c}, E_{2c}, E_{12}$ are coupling energies. Due to the DC superconducting quantum interference device (DC-SQUID) \cite{tinkham2004introduction}, $E_{Jk}$ can be manipulated with external flux biases through the SQUID loops, $\hat{\phi}_k$ now corresponds to effective phase operators on account of external fluxes. We refer to this general Hamiltonian as our initial model, any QCQ architecture described by \eqref{ham0} applies to the results in this work. Obviously, the second term of the Hamiltonian corresponds to the JJ-layout, while the m-layout is described by the rest terms.

After the second quantization of \eqref{ham0} \cite{krantz2019quantum}, the characteristic parameters such as frequencies $\omega_k=\sqrt{8E_{Ck}E_{Jk}}-E_{Ck}$ and couplings between every two elements (see \ref{section2}) are obtained. It is important to note that $\omega_k$ is thus tunable due to the manipulation on $E_{Jk}$. To ensure the scalability of the QCQ architecture in superconducting quantum chips and for simplicity as well, two qubits (i.e., Q1 and Q2 in Figure~\ref{circuits}(a)) are assumed to be identical, and symmetric about the intermediate coupler, thus we take $ E_{C1}=E_{C2}=E_{Cq}, E_{J1}=E_{J2}=E_{Jq}, \vert E_{1c}\vert =\vert E_{2c}\vert $ and the qubit frequency $\omega_1=\omega_2=\omega_q$. 
Considering dispersive coupling between the qubit and coupler so that the coupler remains in the ground state all the time, and applying a second-order Schrieffer-Wolff transformation, we are able to obtain the QQe coupling strength  \cite{yanTunableCouplingScheme2018}, which reads
\begin{equation}
\begin{aligned}
    g=&& \frac{\omega_q^{lin}}{4}\left(\frac{E_{12}}{E_{Cq}}-\frac{E_{1c}E_{2c}}{2E_{Cq}E_{Cc}}\frac{\omega_c^{lin}\omega_c}{\omega_c^2-\omega_q^2}\right)\\
    \approx&& \frac{\omega_q}{4}\left(\frac{E_{12}}{E_{Cq}}-\frac{E_{1c}E_{2c}}{2E_{Cq}E_{Cc}}\frac{\omega_c^2}{\omega_c^2-\omega_q^2}\right) \label{geff1},
\end{aligned}
\end{equation}
where $\omega_k^{lin}=\sqrt{8E_{Ck}E_{Jk}}$ $(k=q,c)$ is the linear frequency of the qubit and coupler. Since $E_{Jk} \gg E_{Ck}$ in transmon qubits, the frequency and the linear frequency will not be distinguished in the regime of interest. The rationality of using this approximation has been verified both theoretically and numerically.  

Note that for different QCQ structure configurations, the charging energies are always positive. However, it does not hold for coupling energies $E_{12},E_{1c},E_{2c}$. In order to achieve the crucial zero-coupling condition (i.e., $g=0$), $E_{12}$ and $E_{1c}E_{2c}/\left(\omega_c^2-\omega_q^2\right)$ must have the same sign. Thus there are two operating regimes: $\omega_c>\omega_q$ with ${\rm sgn}(E_{12}E_{1c}E_{2c})>0$ and $\omega_c<\omega_q$ with ${\rm sgn} (E_{12}E_{1c}E_{2c})<0$, where ${\rm sgn}$ represents the sign function. 
Although several works have realized two-qubit gates in $\omega_c<\omega_q$ regime \cite{seteFloatingTunableCoupler2021,stehlikTunableCouplingArchitecture2021,marxerLongdistanceTransmonCoupler2022}, the QCQ system in this regime is quite fragile. The lower coupler frequency results in the risk of energy leakage from the computational states of qubits into higher energy states of the coupler. As a consequence, the tunable range of the coupler frequency is restricted considerably, bringing about more complexity in both design and manipulation procedures. Therefore, we will concentrate on the regime $\omega_c>\omega_q$.

Under the regime of interest, the QQe coupling (\ref{geff1}) can be rewritten in a simple form
\begin{eqnarray}
    && g=\frac{2\omega_q}{B}\left(A-\frac{\omega_c^2}{\omega_c^2-\omega_q^2}\right)\label{geff2} , \\
    && A = \frac{2E_{12}E_{Cc}}{E_{1c}E_{2c}},\quad B = \frac{16E_{Cq}E_{Cc}}{E_{1c}E_{2c}}
\end{eqnarray}
where $A, B$ are both dimensionless variables.
The previous discussion indicates that $A$ keeps positive all the time while $B$ shares the same sign of $E_{12}$. Because both of the charging energies and coupling energies are determined only by the m-layout which can be abstracted as a set of capacitances $\mathcal{C}$ (including self capacitances of each element and mutual capacitances among different elements), $A=A(\mathcal{C})$ and  $B=B(\mathcal{C})$ play an important role in QQe coupling characteristics.

For later convenience, we introduce an important quantity: the dispersive rate of qubit-coupler coupling $\beta$ that is defined as
\begin{equation}\label{dispersive rate}
    \frac{1}{\beta} \equiv \frac{\left\vert g_{qc}\right\vert }{\omega_c-\omega_q} \approx \frac{1}{\sqrt{\vert B\vert }}\frac{\sqrt{\omega_q\omega_c}}{\omega_c-\omega_q},
\end{equation}
where $g_{qc}$ is the qubit-coupler coupling strength since two qubits are identical. The dispersive coupling requires $\beta \gg 1$ which is indeed the prerequisite of \eqref{geff1} as well.  In the layout design or manipulation procedure, we generally set up a lower limit of the dispersive rate $\beta_s$ above which the energy exchange between the qubits and the coupler is suppressed largely and thus negligible. 

\section{Results and discussions}

\subsection{Zero-coupling condition} \label{subsec.zero}

The zero-coupling condition can be satisfied, i.e. the solution to $g = 0$ exits, under the condition
\begin{equation}\label{zero coupling condition} 
    1 < A \le \frac{1}{2}\left(1+\sqrt{1+4 \vert   B \vert  /\beta_s^2}\right).
\end{equation}
The detailed derivations can be found in  \ref{section2}. Note that $A>1$ is always a necessary condition even when the assumption of identical and symmetric qubits breaks down.
Since $\beta_s$ is given in advance and $A, B$ are determined by $\cal{C}$, this condition only depends on the m-layout. As an essential application, one can either decide in advance which QCQ configuration is appropriate or estimate whether or not a QCQ  m-layout is qualified after the design. For instance,  
once we intend to realize long-range qubit-qubit couplings in QCQ architectures, the condition (\ref{zero coupling condition}) immediately tells us that choosing the grounded type for the coupler is no longer appropriate. More precisely, taking G-G-G configuration shown in Figure~\ref{circuits}(b) for instance, $A$ reads (see the  derivations in \ref{section3})
$$
    A_{\text{G-G-G}} = \frac{(C_{0q}+C_{qc})(C_{qc}^2+C_{0c}C_{qq}+2C_{qc}C_{qq})}{C_{qc}^2(C_{0q}+C_{qc}+2C_{qq})}
$$ 
where $C_{0q}, C_{0c}$ are the self capacitances of the qubit and coupler, $C_{qc}, C_{qq}$ are the qubit-coupler, qubit-qubit mutual capacitances, respectively. The long-range qubit-qubit coupling means the direct mutual capacitance between qubits nearly vanishes, namely $C_{qq}=0$. Therefore, we get $A_{\text{G-G-G}}=1$, which implies the long-range QCQ structure with G-G-G configuration can never achieve $g=0$ no matter how well the other capacitances and JJ-layouts are managed. This conclusion indeed also holds for the F-G-F configuration, sentencing the grounded-type coupler of no use in long-range qubit-qubit coupling architecture. More detailed derivation can be found in \ref{section3}.


\subsection{Upper bound of QQe coupling}

When the QQe coupling is turned on, the coupling strength is generally expected to be stronger for implementing faster two-qubit gates. However, several restraints prevent its realization. In addition to the zero-coupling condition and the dispersive limit $\beta \ge \beta_s$, one more restraint is the limited element frequency due to possible restrictions in chip fabrication, manipulation, and related facilities. Therefore, a possible upper limit on the coupler frequency $\omega_l$ is considered. 

In practice,
we mainly concern two critical coupler frequencies $\omega_{\text{on}}, \omega_{\text{off}}$, where the QQe coupling is turned on or off respectively. The QQe coupling in the $\omega_c>\omega_q$ regime also has two different sub-regimes dependent on the arrangement of $\omega_{\text{on}}, \omega_{\text{off}}$ as shown in Figure~\ref{circuits}(c), where $\omega_s, \omega_l$ represent the smaller one and the larger one of the critical frequencies. By this definition, the qubit-coupler dispersive rate reaches the lower limit $\beta_s$ at $\omega_s$.
Fortunately, these two sub-regimes can be analyzed uniformly if we consider the absolute value of QQe coupling strength at $\omega_{\text{on}}$. By introducing  two additional dimensionless variables 
$x = \omega_q/\omega_s, x \in (0,1)$ and $y = \omega_s/\omega_l, y \in (0,1)$ as an essential technique, we are able to extract the restraints $\beta_s,\omega_l$ as detached pre-factors, leading to a concise presentation of QQe coupling (see the detailed derivations in \ref{section4}) 
\begin{equation}\label{gon}
    \left\vert g_{\text{on}}\right\vert  = \frac{2 \omega _l}{\beta _s^2}\left[\frac{(1-x)  \left(1-y^2\right)x^2 y}{(1+x) \left(1-x^2 y^2\right)}\right]
    =\frac{2 \omega _l}{\beta _s^2}f(x,y).
\end{equation}
All the other quantities of interest can also be depicted with $\omega_l, \beta_s, x, y$. In particular, $A$ is different while $B, \omega_q, \omega_s$ are the same in two sub-regimes, they read
\begin{equation}\label{other quantities}
\left\{
\begin{aligned}
    A = &\left\{
                    \begin{aligned}
                        \frac{1}{1-x^2}, & \quad \omega_{\text{on}} > \omega_{\text{off}}\\
                        \frac{1}{1-x^2y^2}, &  \quad \omega_{\text{on}} < \omega_{\text{off}}
                    \end{aligned}
                    \right. \\
    \vert B\vert  = &\frac{x}{(1-x)^2}\beta_s^2 \\
    \omega_q =& xy\omega_l, \quad \omega_s = y\omega_l
\end{aligned}
\right..
\end{equation}


For fixed $\omega_l$ and $\beta_s$, we are able to obtain the maximum $\left\vert g_{\text{on}}\right\vert _{max}$. An analytical or numerical analysis of $f(x,y)$ results in a maximum value $0.0411$ at $x\approx 0.675, y\approx 0.652$ as shown in Figure~\ref{f(x,y)}, which leads to an upper bound of the QQe coupling strength
\begin{equation}\label{geffmax}
    \left\vert g_{\text{on}}\right\vert _{max} = 0.0822\frac{\omega _l}{\beta _s^2}.
\end{equation}
Significantly and surprisingly, this bound only depends on the pre-decided quantities $\omega_l$ and $\beta_s$. This robust conclusion reveals the extremity of the two-qubit gate performance in QCQ architecture from a global perspective. One can immediately estimate the maximum QQe coupling with artificially given  $\omega_l$ and $\beta_s$, totally independent of any detailed layout design.
Taking $\omega_l=15$ GHz and $\beta_s=10$ for instance, using (\ref{geffmax}) the maximum achievable QQe coupling is evaluated as $12.33$ MHz. In addition, using $\omega_q = xy\omega_l$, (\ref{geffmax}) can be rewritten as $\left\vert g_{\text{on}}\right\vert _{max} = 0.187~\omega _q/\beta _s^2$. The upper bound can be used to explain roughly a part of experimental data in recent related work \cite{aruteQuantumSupremacyUsing2019,wuStrongQuantumComputational2021,zhuQuantumComputationalAdvantage2022} as summarized in Table~\ref{data} where we assume $\beta_s=8$ approximately, despite that the detailed information of the layouts and coupler frequencies is unknown.

\begin{figure}[htbp]
\centering
\includegraphics[width=0.6\linewidth]{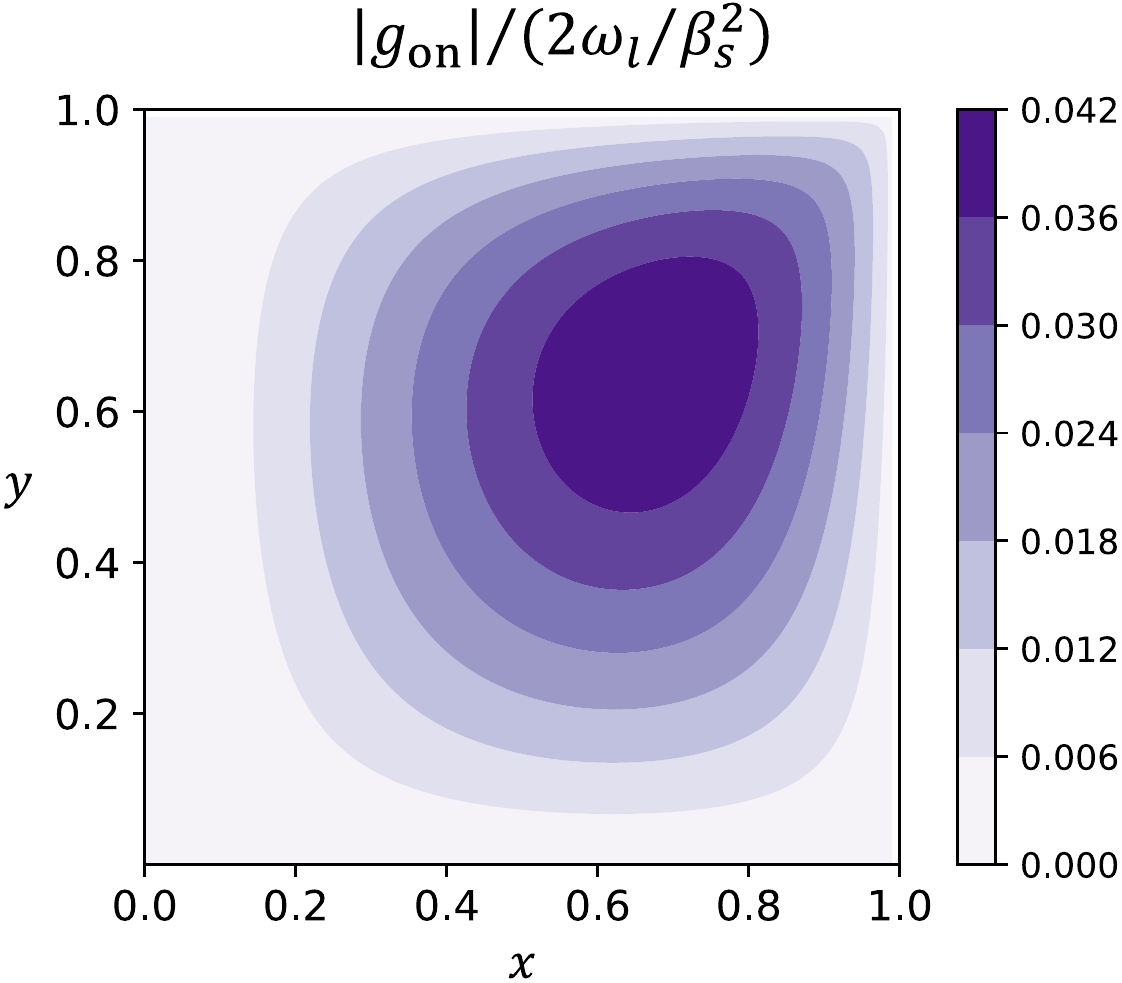}
\caption{ (Color online) The QQe coupling $\left\vert g_{\text{on}}\right\vert $ characteristics dependent of $x$ and $y$. Apparently, a maximum QQe coupling does exist in QCQ architecture.}
\label{f(x,y)}
\end{figure}

\begin{table}[htbp]
\caption{\label{data}A part of experimental Data for different existing quantum processors, including \textit{Zuchongzhi 2.0} \cite{wuStrongQuantumComputational2021}, \textit{Zuchongzhi 2.1} \cite{zhuQuantumComputationalAdvantage2022}, and \textit{Sycamore} \cite{aruteQuantumSupremacyUsing2019}. The reference $\omega_q$ refers to the median value among all qubits at their maximum frequencies, and the reference $\left\vert g_{\text{on}}\right\vert $ refers to the coupling strength used to implement two-qubit gates. The $\left\vert g_{\text{on}}\right\vert _{max}$ is evaluated using $0.187~\omega_q/\beta_s^2$ where we take reference $\omega_q$ and  $\beta_s=8$. Our estimated results exhibit high consistency with reference data.}
\begin{indented}
\item[]
\begin{tabular}{@{}llll@{}}
\br
  Processor & \textit{Zuchongzhi 2.0} & \textit{Zuchongzhi 2.1}  & \textit{Sycamore}  \\
\mr
    reference $\omega_q$ & 4.811 GHz & 5.099 GHz & 6.924 GHz \\
    reference $\left\vert g_{\text{on}}\right\vert $ & $\sim 10$ MHz & $\sim 14$ MHz & $\sim 20$ MHz \\
    \textbf{estimated} $\bm{\left\vert g_{\mathrm{on}}\right\vert _{max}}$ & \textbf{14.06 MHz} & \textbf{14.90 MHz} & \textbf{20.23 MHz} \\
\br
\end{tabular}
\end{indented}
\end{table}

\subsection{Optimal design procedure}

Nevertheless, the layout can be subtly designed to reach the upper bound $\left\vert g_{\text{on}}\right\vert _{max}$. As discussed, the m-layout is mainly determined by $A$ and $B$, therefore the optimal m-layout emerges in researching optimal $A,B$. As a contrast, the JJ-layout depends on the related optimal frequencies $\omega_q$ and $\omega_l$. More specifically,  the optimal $A, B$ can be calculated using the relations between $A, B$ and $x,y$ (the optimal  $x$ and $y$ are  0.675,  0.652 respectively) obtained before, leading to a set of capacitance equations:
\begin{equation}\label{bestAB}
\left\{
\begin{aligned}
    A(\mathcal{C}) \approx &\left\{
                    \begin{array}{cc}
                        1.84, & ~~~~~  \quad \omega_{\text{on}} > \omega_{\text{off}}\\
                        1.24, & ~~~~~ \quad \omega_{\text{on}} < \omega_{\text{off}}
                    \end{array}
                    \right. \\
    B(\mathcal{C}) \approx &\left\{
                    \begin{array}{cc}
                        6.4 ~ \beta_s^2, &  \quad E_{12}>0\\
                        -6.4 ~ \beta_s^2, &  \quad E_{12}<0
                    \end{array}
                    \right.
\end{aligned}
\right..
\end{equation}
The sign of $E_{12}(\mathcal{C})$ can be recognized in simple cases such as the G-G-G configuration, where one can always manage $E_{12}$ to be positive or negative. However, the situation is different in some complicated cases where the sign $E_{12}(\mathcal{C})$ is hard to tell ahead of time, thus we need to take both cases of $B$ for discussion instead. More discussions concerning this issue can be seen in \ref{section5}.
To settle down an optimal m-layout, in addition to the above two equations (i.e., \eqref{bestAB}) the other $n(\mathcal{C})-2$ equations of capacitances are needed, where $n(\mathcal{C})$ is the cardinal number of $\mathcal{C}$. This implies that the optimal m-layout is not unique, providing enough degree of freedom and significant flexibility for us to count in other possible restraints.
Besides, the optimal qubit frequency and the lower critical frequency of the coupler are evaluated as
$\omega_q \approx 0.44 ~ \omega_l \label{optimal wq}, \omega_s \approx 0.652 ~ \omega_l \label{optimal ws}$.

\begin{figure}[htbp]
\centering
\includegraphics[width=0.6\linewidth]{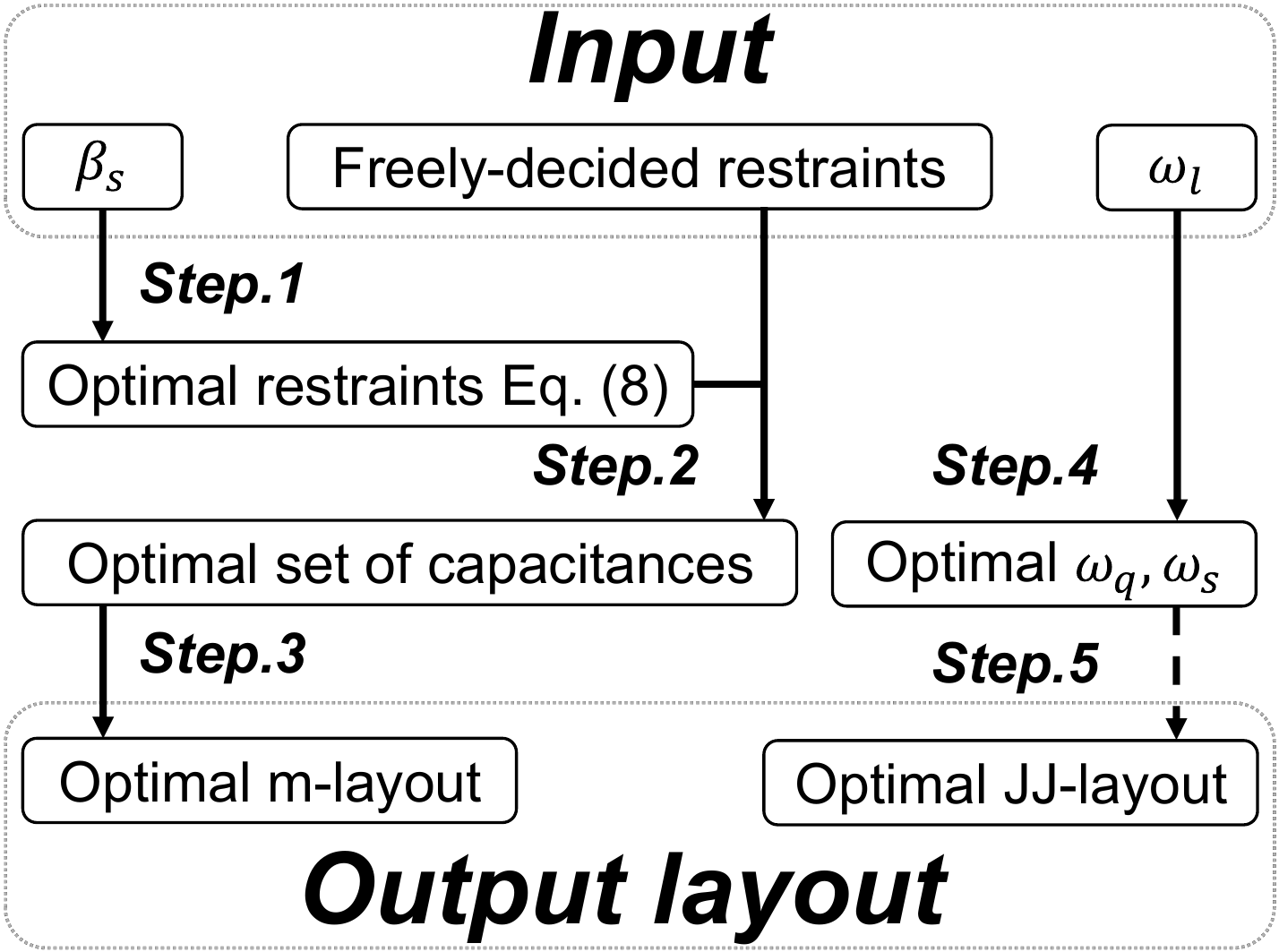}
\caption{The optimal layout design procedure. Following this workflow, the layout of coupler architecture can be designed automatically and efficiently. }
\label{procedure}
\end{figure}

As a crucial conclusion, we offer a systematic procedure for optimal layout design shown in Figure~\ref{procedure}.  One remarkable feature is that the m-layout and JJ-layout can be designed independently. In the next section, we will exactly follow this workflow to design specific layouts. 

\begin{figure*}[htbp]\centering
\includegraphics[width=\linewidth]{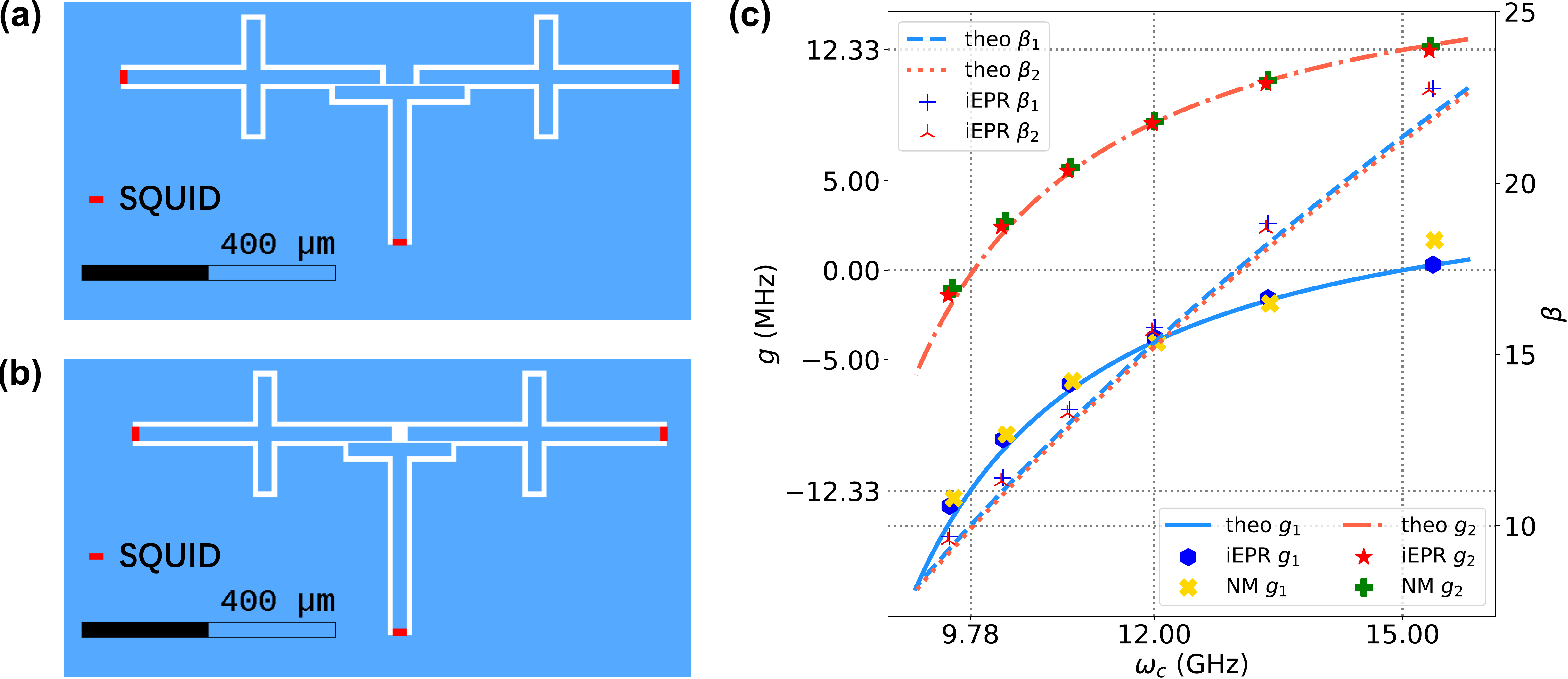}
\caption{ (Color online) (a) and (b) are the QCQ layouts corresponding to $\omega_{\text{on}}<\omega_{\text{off}}$ and $\omega_{\text{on}}>\omega_{\text{off}}$ regimes respectively. The regions in blue are the superconducting metal while the little red rectangles represent the SQUIDs which are treated as lumped inductances in electromagnetic simulation. 
Two qubits are located on the left and right sides with a tunable coupler capacitively coupled in the middle. 
We use sapphire as the substrate whose relative permittivity is 10. 
(c) The QQe coupling $g$ and qubit-coupler dispersive rate $\beta$ characteristics tuned by the coupler frequency $\omega_c$ in two different sub-regimes. It contains both theoretical (theo) and electromagnetic simulation experimental results, where the subscripts $1,2$ refer to $\omega_{\text{on}}<\omega_{\text{off}}$ and $\omega_{\text{on}}>\omega_{\text{off}}$ regimes respectively. The QQe coupling $g$ is verified both by iEPR and normal-mode (NM) techniques while the dispersive rate $\beta$ is benchmarked by iEPR technique.}
\label{layout}
\end{figure*}

\section{Verification and Applications}

\subsection{Optimal QCQ layout design}

To demonstrate the validity and high efficiency of the design procedure, we apply it to design an optimal layout with G-G-G configuration. Following the workflow illustrated in Figure~\ref{procedure} processively, without loss of generality we take $\beta_s=10$ in Step.1, and obtain $A(\mathcal{C})=1.84~ (1.24), B(\mathcal{C})=640$ for $\omega_{\text{on}} > \omega_{\text{off}}~ (\omega_{\text{on}} < \omega_{\text{off}})$ regime where we managed $E_{12}$ to be positive. The explicit expressions in terms of capacitances for $A(\mathcal{C})$ and $B(\mathcal{C})$ are derived using cQED theory, and we presented them in \ref{section3}. In Step.2, we first take a desired qubit anharmonicity, e.g., $-E_{Cq}(\mathcal{C})=-230$ MHz, and a certain mutual capacitance, e.g., $C_{qc}=6$ fF, for extra restraints. Then by solving these four capacitance equations simultaneously, we obtain the optimal sets of capacitances for two different sub-regimes, as shown in Table~\ref{layoutdata}. In step.3, the m-layouts corresponding to two different sub-regimes are procured with aid of capacitance simulation, as shown in Figure~\ref{layout}(a) and (b) respectively. In Step.4, we take $\omega_l=15$ GHz in both regimes, leading to a 12.33 MHz maximum QQe coupling. Furthermore, the corresponding optimal $\omega_q$ and $\omega_s$ can also be calculated, and the results are also presented in Table~\ref{layoutdata}. The technique details for Step.5 are omitted since JJ-layouts are indeed related highly to the specific design and fabrication restraints. Till now, we have completed the whole layout design.

A natural task followed is to examine the characteristics of the layouts. The key characterized parameters $g$ and $\beta$ for two sub-regimes are calculated theoretically using \eqref{geff2} and \eqref{dispersive rate}, and verified through 
 performing electromagnetic simulation experiments scrupulously based on two different methods respectively: i) the existing normal mode (NM) analysis technique \cite{dubyna_inter-qubit_2020} and ii) iEPR technique (developed by K.Y., L.J., et al., the manuscript is in preparation), an outstanding extension of the energy-participation ratio (EPR) method \cite{minevEnergyparticipationQuantizationJosephson2021a} based on electromagnetic eigenmode simulation. As shown in Figure~\ref{layout}(c), the QQe couplings across the zero-coupling line at $\omega_{\rm off}$ and achieve a stronger QQe coupling at $\omega_{\rm on}$, moreover maintain dispersive couplings all the time. Note that $\beta$ is taken as the average value of two identical-qubit-coupler dispersive rates since they are the same in theory but slightly diverge in electromagnetic simulation experiments. In addition, all the elements are treated as linear harmonic oscillators here for an equitable comparison of theoretical and simulation experimental results due to the linear characteristic in electromagnetic simulation. Furthermore, the results obtained by theory and electromagnetic simulation experiments are highly consistent, revealing the effectiveness of the layout design procedure. Besides, the slight geometric differences between the layouts (i.e., Figure~\ref{layout}(a) and Figure~\ref{layout}(b))  indeed cause a significant distinction in performance, revealing the hardness and vulnerability of the practical layout design. Therefore, the procedure proposed in this work provides great convenience, efficiency, and performance superiority for the practical layout design.

\begin{table}[htbp]
\setlength{\tabcolsep}{1pt}
\caption{\label{layoutdata}The detailed information for the two different designed layouts containing the sets of capacitances for the optimal m-layouts, the optimal  characterized frequencies and maximum QQe couplings with $\omega_l=15$ GHz.}
\begin{indented}
\item[]    
        \begin{tabular}{@{}l|llllllll@{}}
        \br
            & $C_{0q}$ (fF) & $C_{0c}$ (fF) & $C_{qc}$ (fF) & $C_{qq}$ (fF) & $\omega_{\text{on}}$ (GHz) & $\omega_{\text{off}}$ (GHz) & $\omega_q$ (GHz) & $g_{\text{on}}$ (MHz) \\
        \mr
           $\omega_{\text{on}} < \omega_{\text{off}}$ & 56.6 & 78.6 & 6 & 0.128 & 9.78 & 15 & 6.61 & -12.33\\
           $\omega_{\text{on}} > \omega_{\text{off}}$ & 57.1 & 78.3 & 6 & 0.448 & 15 & 9.78 & 6.61 & 12.33 \\
        \br
        \end{tabular} 
\end{indented}
\end{table}

\subsection{Long-range QCQ layout design}

As the validity of the layout design procedure is well verified, a more conspicuous application is designing a long-range QCQ layout. The strong motivation behind is as follows: 
i) opening up more space for the individual readout resonator and Purcell filter of each qubit leading to fast and high-fidelity readout \cite{heinsooRapidHighfidelityMultiplexed2018,seteQuantumTheoryBandpass2015}; 
ii) enabling a larger number of qubits integrated in superconducting quantum chips because wiring  problems are improved dramatically;
iii) mitigating unwanted non-nearest-neighbor coupling and possible crosstalk;
iv)  suppressing correlated errors which is crucial to quantum error correction \cite{wilenCorrelatedChargeNoise2021}.

As discussed in \ref{subsec.zero}, a floating coupler is required in long-range qubit-qubit coupling layouts. Actually, the floating coupler architecture was first proposed in  \cite{seteFloatingTunableCoupler2021} to realize long-range qubit-qubit effective coupling experimentally. Very recently, \cite{marxerLongdistanceTransmonCoupler2022} reaches a qubit-qubit distance of 1960 um far apart by introducing a waveguide extender between the qubit and coupler. This way, they subtly get rid of the complexity and difficulty in designing long-range and high-performance QCQ layouts with direct qubit-coupler mutual capacitances. In this work, however, we demonstrate that with aid of the key findings, we are able to design a practicable QCQ layout of longer qubit-qubit distance and higher performance with the original floating coupler architecture.

\begin{figure*}[htbp]\centering
\includegraphics[width=\linewidth]{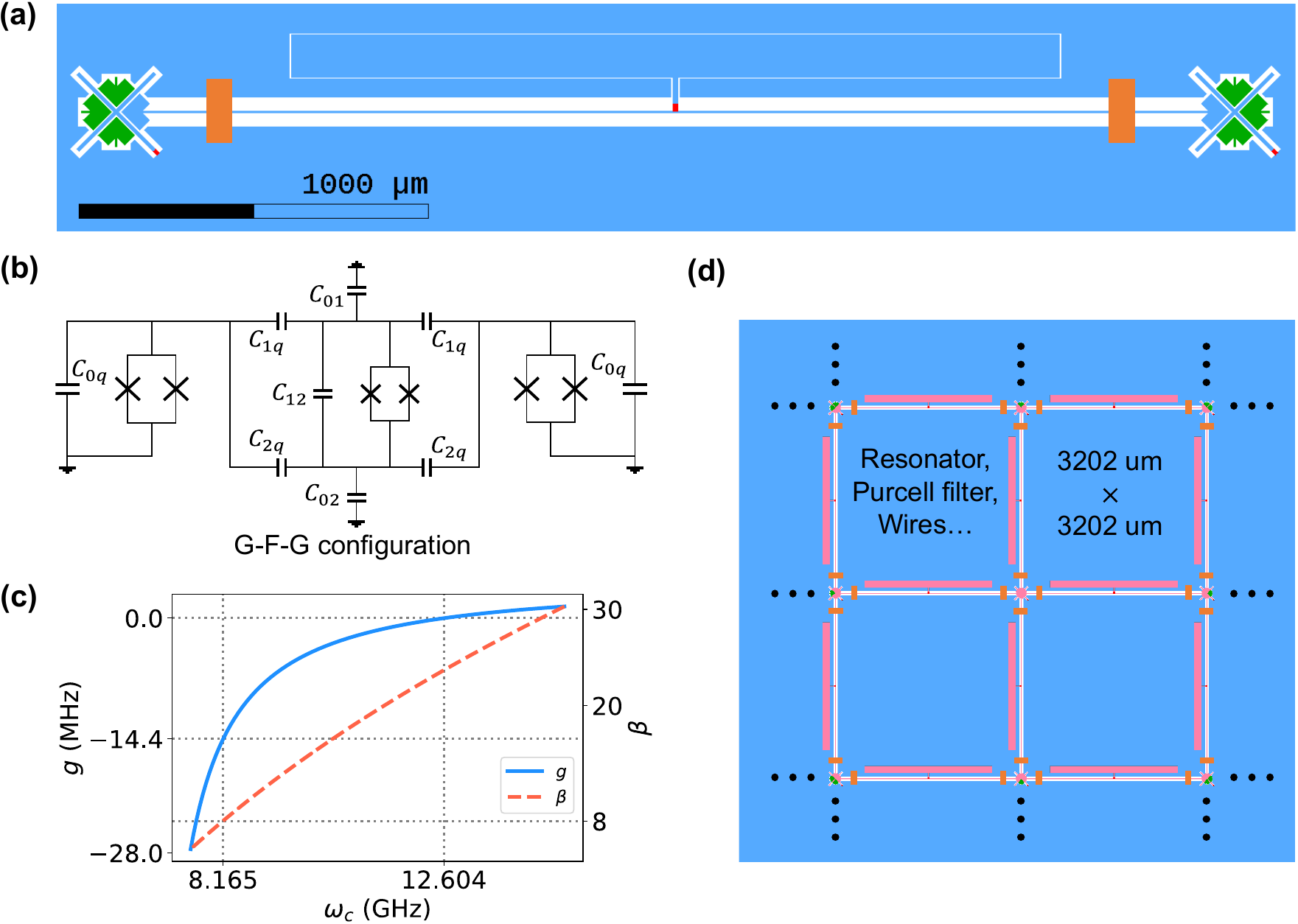}
\caption{(Color online) (a) The layout of the scalable long-range QCQ module. Two grounded qubits are placed far apart with a floating coupler capacitively coupled in the middle. The tiny rectangles (red) represent SQUIDs, and the rectangles (orange) are airbridges whose function is to balance the potential on both sides and make the ground complete. The pads (green) around qubits are the alternative parts used to scale this QCQ module. The substrate is taken with sapphire of relative permittivity 10, and the flip-chip architecture is considered where two opposite chips are 10 um apart. (b) The equivalent circuit diagram of the QCQ system in G-F-G configuration. (c) The expected QQe coupling $g$  and dispersive rate $\beta$ characteristics. (d) The scalable square lattices using the long-range QCQ module, opening up enough space for handling the tough wiring, crosstalk, correlated errors problems. The qubits and couplers are re-colored in pink for highlighting.
}
\label{gfg layout}
\end{figure*}

In particular, we take the G-F-G configuration in $\omega_{\text{on}} < \omega_{\text{off}}$ regime and consider $\beta_s=8$. The equivalent circuit is shown in Figure~\ref{gfg layout}(b), where we consider the mutual capacitance between two qubits to be zero due to the long-range coupling, leading to $\mathcal{C}=\{C_{0q},C_{01},C_{02},C_{12},C_{1q},C_{2q}\}$. As a rule of thumb, however, we are not able to reach the optimal layout in practice due to possible limitations of layout geometry on the capacitance values. Nevertheless, we can adjust $x$ and $y$ slightly off the optimal values to realize a practicable layout while simultaneously reserving high performance as far as possible. As an instance, we take $x=0.8$ and $ y=0.652$ in our design. Using \eqref{other quantities}, we obtain $A(\mathcal{C})=1.37, B(\mathcal{C})=-1280$ (where $E_{12}$ is negative). Next step, we take $E_{Cq}(\mathcal{C}),C_{12},C_{1q},C_{2q}$ as four other freely decided restraints and adjust their values to realize a practicable layout, finally leading to a long-range QCQ layout shown in Figure~\ref{gfg layout}(a) where the distance between two qubits is up to 3202 um. The capacitances corresponding to the actual layout are shown in Table~\ref{gfg data} and the qubit anharmonicity is -236 MHz. In absence of the well-defined design procedure introduced in this work, it would result in great difficulty due to the large number of  
capacitance parameters involved.
Next, we set $\omega_q=6.5$ GHz, according to \eqref{gon} and \eqref{other quantities}, we can switch off the QQe coupling at $\omega_l\approx 12.6$ GHz and realize a strong QQe coupling of $14.4$ MHz at $\omega_s\approx 8.2$ GHz where the dispersive rate reaches its lower limit $\beta_s$. The predicted performance using \eqref{geff2} and \eqref{dispersive rate} is shown in Figure~\ref{gfg layout}(c).

Another remarkable feature is the significant scalability. Making use of the additional coupling pads around qubits whose structure is the same as the ends of the coupler, we can easily scale up this QCQ module to square lattices by replacing coupling pads with couplers as shown in Figure~\ref{gfg layout}(d).
Consequently, each qubit is enabled to own an individual readout resonator and Purcell filter in the large 3202 um $\times$ 3202 um square lattice for faster high-fidelity readout, and the qubits are far apart enough to better get rid of correlated errors and non-nearest-neighbor coupling. More importantly, the quantum error correction with surface code could be significantly enhanced due to the high scalability of this layout \cite{acharyaSuppressingQuantumErrors2022}. 

\begin{table}[htbp]
\caption{\label{gfg data}The capacitance information of the long-range QCQ layout.}
\begin{indented}
\item[]
        \begin{tabular}{@{}llllll@{}}
        \br
             $C_{12}$ (fF) & $C_{1q}$ (fF) & $C_{2q}$ (fF) & $C_{0q}$ (fF) & $C_{01}$ (fF) & $C_{02}$ (fF) \\
        \mr 
           11.8 & 10.4 & 0.04 & 71.9 & 266.3 & 840.1 \\
        \br
        \end{tabular} 
\end{indented}
\end{table}

\section{Summary and perspectives}

In summary, we applied cQED theory to the QCQ architecture from the layout perspective and came up with three crucial results: a zero-coupling condition (\eqref{zero coupling condition}), an upper bound of QQe coupling (\eqref{geffmax}), and an optimal layout design procedure (Figure~\ref{procedure}). 
Among these findings, we want to highlight especially the QQe coupling extremity, namely $0.0822~ {\omega _l}/{\beta _s^2}$. Its dependence on the artificially given quantities $\omega_l, \beta_s$ instead of specific layouts makes it a robust conclusion from a global architecture view and therefore significantly reveals the limitation of the prevalent tunable coupler architecture. 
In addition, the well-defined layout design procedure is expected to be used as a prominent instruction through which anyone can be an expert in designing QCQ layouts of high performance. Last but not least, if one has special requirements and the procedure is not essential, the zero-coupling condition could kindly offer some help before or after the layout design. As a stirring application, we report a state-of-the-art and scalable long-range QCQ layout, providing a promising solution for superconducting quantum chips with larger scales and higher performance.
Although we fill up the gap between the design demands and capacitance parameters, how to transform automatically the targeted capacitance parameters to the geometric parameters remains to be studied.
Looking forward, the train of thought and techniques introduced in this work could be extended to other architectures or other components, which will no doubt accelerate the research and development of superconducting quantum chips. 
We also expect more novel and promising QCQ layouts with different distinguishing features will be developed in an efficient way, and the superconducting quantum chip design automation will be moved forward to a new level as well.

\section*{Data Availability}
The data that support the findings of this work are available upon reasonable
request.

\section*{Acknowledgement}

We would like to thank Yu-Xuan Wang and Ke-Hui Yu for helpful discussions.
This work was done when F.L. was research interns at Baidu Research.

\appendix

\section{Derivation of charging and coupling energies in terms of capacitances}
\label{section1}

In the main text, we mentioned that charging energies $E_{C1},E_{C2},E_{Cc}$ and capacitive coupling energies $E_{1c}, E_{2c}, E_{12}$ are all functions of $\mathcal{C}$, here we show the derived procedure to obtain the explicit form. Let us start from the QCQ system Lagrangian:
\begin{eqnarray}
    \mathcal{L} &=& T - U,\\
    T &=& \frac{1}{2}\dot{\mathbf{\Phi}}^T_M\mathbf{M}\dot{\mathbf{\Phi}}_M, \\
    U &=& \sum_{k\in\{q_1,q_2,c\}}E_{Jk}\left(1-\cos\left(\frac{2\pi\Phi_{\Delta_k}}{\Phi_0}\right)\right),
\end{eqnarray}
where $T$ and $U$ are the kinetic and potential energy respectively; $\dot{\mathbf{\Phi}}_M=\text{vec}(\dot{\Phi}_{1}, \dot{\Phi}_{2}, ...)^T$ where $\Phi_\mu$ $(\mu=1,2,...)$ is the node flux of the $\mu$-th metal plate and $\dot{\Phi}_\mu$ corresponds to its node potential; $\mathbf{M}$ is the Maxwell capacitance matrix where the diagonal element $\mathbf{M}_{\mu\mu}$ is the sum of self-capacitance and all the mutual capacitances for the $\mu$-th metal plate, and the off-diagonal element $\mathbf{M}_{\mu\nu}$ $(\mu,\nu=1,2,...,\mu\neq\nu)$ is the negative value of the mutual capacitance between the $\mu$-th and $\nu$-th metal plates; $q_1,q_2,c$ refer to two qubits and the coupler respectively; $\Phi_{\Delta_k}$ is the flux through the element's Josephson junction (or SQUID), which we call the junction flux. Rather than the node flux $\Phi_\mu$, the junction flux $\Phi_{\Delta_k}$ is the one who contributes to the energy oscillation. Therefore, we will use junction fluxes as the primary variables in the Hamiltonian. In addition, to be able to distinguish elements and metal plates clearly,  it is noted that we use $q_1,q_2,c$ as subscripts of two qubits and the coupler respectively in Sec.~\ref{section1},\ref{section3},\ref{section5} instead of $1,2,c$ in the main text.

For grounded elements which contain single metal plates, the junction flux $\Phi_{\Delta_k}$ equal to the node flux $\Phi_k$ naturally. In contrast, for floating elements which contain double metal plates, the junction flux reads $\Phi_{\Delta_k}=\Phi_{k1}-\Phi_{k2}$ where $\Phi_{k1}, \Phi_{k2}$ represent the node fluxes of two metal plates. For substitution, we introduce an auxiliary flux $\Phi_{\Sigma_k}=\Phi_{k1}+\Phi_{k2}$ \cite{seteFloatingTunableCoupler2021}, leading to 
\begin{equation}
    \begin{pmatrix}
         \Phi_{k1} \\
         \Phi_{k2}
    \end{pmatrix}
    = \begin{pmatrix}
         {1}/{2} & {1}/{2} \\
         {1}/{2} & -{1}/{2}
    \end{pmatrix}
    \begin{pmatrix}
         \Phi_{\Sigma_k} \\
         \Phi_{\Delta_k}
    \end{pmatrix}.
\end{equation}
Together with $\Phi_k=\Phi_{\Delta_k}$ for grounded elements, we can build a transform matrix $\mathbf{S}$ for the transformation between node fluxes and junction fluxes. Taking grounded qubits $q_1,q_2$ and floating coupler $c$ for instance,
the transformation reads $\mathbf{\dot{\Phi}}_M =\mathbf{S} \mathbf{\dot{\Phi}}$ where
\begin{equation}
    \dot{\mathbf{\Phi}}_M=
    \begin{pmatrix}
    \dot{\Phi}_{q_1} \\
    \dot{\Phi}_{c1} \\
    \dot{\Phi}_{c2}\\
    \dot{\Phi_{q_2}}
    \end{pmatrix}, 
    \dot{\mathbf{\Phi}}=
    \begin{pmatrix}
    \dot{\Phi}_{\Delta_{q_1}} \\
    \dot{\Phi}_{\Sigma_{c}} \\
    \dot{\Phi}_{\Delta_{c}} \\
    \dot{\Phi}_{\Delta_{q_2}}
    \end{pmatrix},  
    \mathbf{S} = \begin{pmatrix}
    1 & 0 & 0 & 0 \\
    0 & {1}/{2} & {1}/{2} & 0 \\
    0 & {1}/{2} & -{1}/{2} & 0 \\
    0 & 0 & 0 & 1
    \end{pmatrix}.
\end{equation}
Note that  the element orders in $\mathbf{\dot{\mathbf{\Phi}}}_M, \mathbf{\dot{\mathbf{\Phi}}}$, and $\mathbf{S}$ must be consistent. Then we can rewrite the kinetic energy as 
\begin{equation}
    T = \frac{1}{2}\dot{\mathbf{\Phi}}^T\mathbf{C}\dot{\mathbf{\Phi}},
\end{equation}
where $\mathbf{C} = \mathbf{S}^T\mathbf{M}\mathbf{S}$ naturally. 

Introducing the charge variables 
\begin{equation}
    Q_{\Delta_k} = \frac{\partial\mathcal{L}}{\partial\dot{\Phi}_{\Delta_k}}, \quad Q_{\Sigma_k} = \frac{\partial\mathcal{L}}{\partial\dot{\Phi}_{\Sigma_k}},
\end{equation}
the kinetic energy is transformed as 
\begin{equation}
    T = \frac{1}{2}\mathbf{Q}^T\mathbf{C}^{-1}\mathbf{Q},
\end{equation}
where $\mathbf{Q}=\text{vec}(Q_{\Delta_{q_1}}, Q_{\Sigma_c}, Q_{\Delta_c}, Q_{\Delta_{q_2}})^T$. Again we introduce new variables, the Cooper pair number operators as $\hat{n}=Q/2e$, the kinetic energy further reads
\begin{equation}
    T = 4e^2 \cdot \frac{1}{2}\mathbf{N}^T\mathbf{C}^{-1}\mathbf{N},
\end{equation}
where $\mathbf{N} = \text{vec}(\hat{n}_{\Delta_{q_1}}, \hat{n}_{\Sigma_c}, \hat{n}_{\Delta_c}, \hat{n}_{\Delta_{q_2}})^T$. As we mentioned before, only $\Phi_{\Delta_k}$ is engaged in the energy oscillation, thus the element mode is represented by $\hat{n}_{\Delta_k}$ \cite{seteFloatingTunableCoupler2021}, namely $\hat{n}_k=\hat{n}_{\Delta_k}, k\in\{q_1,q_2,c\}$ for both grounded and floating elements. By expanding $T$ and omitting all the  $\hat{n}_{\Sigma_k}$ terms, we obtain 
\begin{equation}
    T = \sum_{k\in\{q_1,q_2,c\}}4E_{Ck}\hat{n}_k^2+\sum_{i,j\in\{q_1,q_2,c\},i\neq j}4E_{ij}\hat{n}_i\hat{n}_j,
\end{equation}
where 
\begin{eqnarray}
E_{Ck} &=& e^2\frac{\mathbf{C}^{-1}_{\Delta_k,\Delta_k}}{2}, \label{ECk}\\ 
E_{ij} &=& e^2\mathbf{C}^{-1}_{\Delta_i,\Delta_j}, \label{Eij} \quad i,j,k\in\{q_1,q_2,c\}, i\neq j.
\end{eqnarray}

As a consequence, the charging energies $E_{Ck}$ and capacitive coupling energies $E_{ij}$ are expressed in terms of capacitances $\mathcal{C}$. Besides, the result also indicates that all the QCQ systems of different configurations lead to a uniform Hamiltonian presentation as we stressed at the beginning of the main text. In the practical design, we firstly obtain the Maxwell capacitance matrix $\mathbf{M}$ of the layout by capacitance simulation. According to different element structure types, we then build the corresponding transform matrix $\mathbf{S}$ where the element and related metal plate orders are consistent with $\mathbf{M}$. At the end, we are able to obtain $\mathbf{C} = \mathbf{S}^T\mathbf{M}\mathbf{S}$, leading to the concerned $E_{Ck}(\mathcal{C})$ and $E_{ij}(\mathcal{C})$.

\section{Derivation of the zero-coupling condition}\label{section2}

In this section, we derive the zero-coupling condition. Starting with the general case, we apply a second-order
Schrieffer-Wolff transformation to the general Hamiltonian of the  superconducting  coupler architecture, and assume the coupler remains in ground state all the time, obtaining the well-known QQe coupling strength \cite{yanTunableCouplingScheme2018} 
\begin{equation}\label{g0}
    g = g_{12}-\frac{g_{1c}g_{2c}}{2}\left(\frac{1}{\Delta_1}+\frac{1}{\Delta_2}+\frac{1}{\Sigma_1}+\frac{1}{\Sigma_1}\right),
\end{equation}
where $g_{12},g_{1c},g_{2c}$ are qubit-qubit direct coupling and qubit-coupler couplings respectively and $\Delta_j = \omega_c-\omega_j, \Sigma_j=\omega_c+\omega_j$ with $j=1,2$ denoted as two qubits. Since the qubits and  coupler are  coupled capacitively, the coupling strengths can be further expressed in terms of charging energies and Josephson energies:
\begin{equation}\label{gE}
\begin{array}{cccc}
     g_{12}&=&\frac{E_{12}}{\sqrt{2}}\left(\frac{E_{J1}}{E_{C1}}\frac{E_{J2}}{E_{C2}}\right)^{\frac{1}{4}},&\\
     g_{jc}&=&\frac{E_{jc}}{\sqrt{2}}\left(\frac{E_{Jj}}{E_{Cj}}\frac{E_{Jc}}{E_{Cc}}\right)^{\frac{1}{4}},&\quad j\in\{1,2\}.
\end{array}
\end{equation}

Substituting (\ref{gE}) into (\ref{g0}) and 
taking $\omega_k \approx \sqrt{8E_{Ck}E_{Jk}}$, $k\in\{1,2,c\}$, we obtain the QQe coupling expressed in terms of the crucial quantity  $A, B$:
\begin{equation}
    g = \frac{2 \sqrt{\omega_1 \omega_2}}{B} \left[A-\frac{\frac{\omega_c^2}{\omega_c^2-\omega_1^2}+\frac{\omega_c^2}{\omega_c^2-\omega_2^2}}{2}\right], 
\end{equation}
where the dimensionless parameters
$A = 2E_{12}E_{Cc}/(E_{1c}E_{2c})$ and $B = 16\sqrt{E_{C_1}E_{C_2}}E_{Cc}/(E_{1c}E_{2c})$. 

To switch off the QQe coupling, namely let $g=0$, therefore  
$
    A =\left({\omega_c^2}/{(\omega_c^2-\omega_1^2)}+{\omega_c^2}/{(\omega_c^2-\omega_2^2)}\right)/2
$.
Since $\omega_c^2/(\omega_c^2-\omega_j^2)>1,\; j\in\{1,2\}$ always holds in the $\omega_c>\omega_j$ regime, it is natural to obtain the necessary condition for zero-coupling
\begin{equation}
    A >1.
\end{equation}

Next, we look into the upper bound of $A$. In particular, we assume two qubits are identical and symmetric, namely $\omega_1=\omega_2=\omega_q,E_{C1}=E_{C2},\vert E_{1c}\vert =\vert E_{2c}\vert $, A reads
\begin{equation}\label{A}
    A = \frac{\omega_c^2}{\omega_c^2-\omega_q^2},
\end{equation}
and the qubit-coupler dispersive rate can be defined as
\begin{equation}\label{beta}
    \frac{1}{\beta} \approx \frac{1}{\sqrt{\left\vert B\right\vert }}\frac{\sqrt{\omega_c\omega_q}}{\omega_c-\omega_q},
\end{equation}
as we discussed in the main text.

Substituting \eqref{beta} into \eqref{A}, we obtain 
\begin{equation}
    A = \frac{1}{2}\left(1+\sqrt{1+\frac{4\vert B\vert }{\beta^2}}\right).
\end{equation}
Since the whole theory only works in dispersive limit $\beta \ge \beta_s$ ($\beta_s$ denotes the minimum dispersive rate permitted), thus we have
\begin{equation}
    A \le \frac{1}{2}\left(1+\sqrt{1+\frac{4\vert B\vert }{\beta_s^2}}\right).
\end{equation}

\section{Derivation of $A$ in G-G-G and F-G-F configurations}
\label{section3}

\subsection{G-G-G configuration}\label{section3_subsection1}
\begin{figure}[htbp]
    \centering
    \includegraphics[width=0.7\linewidth]{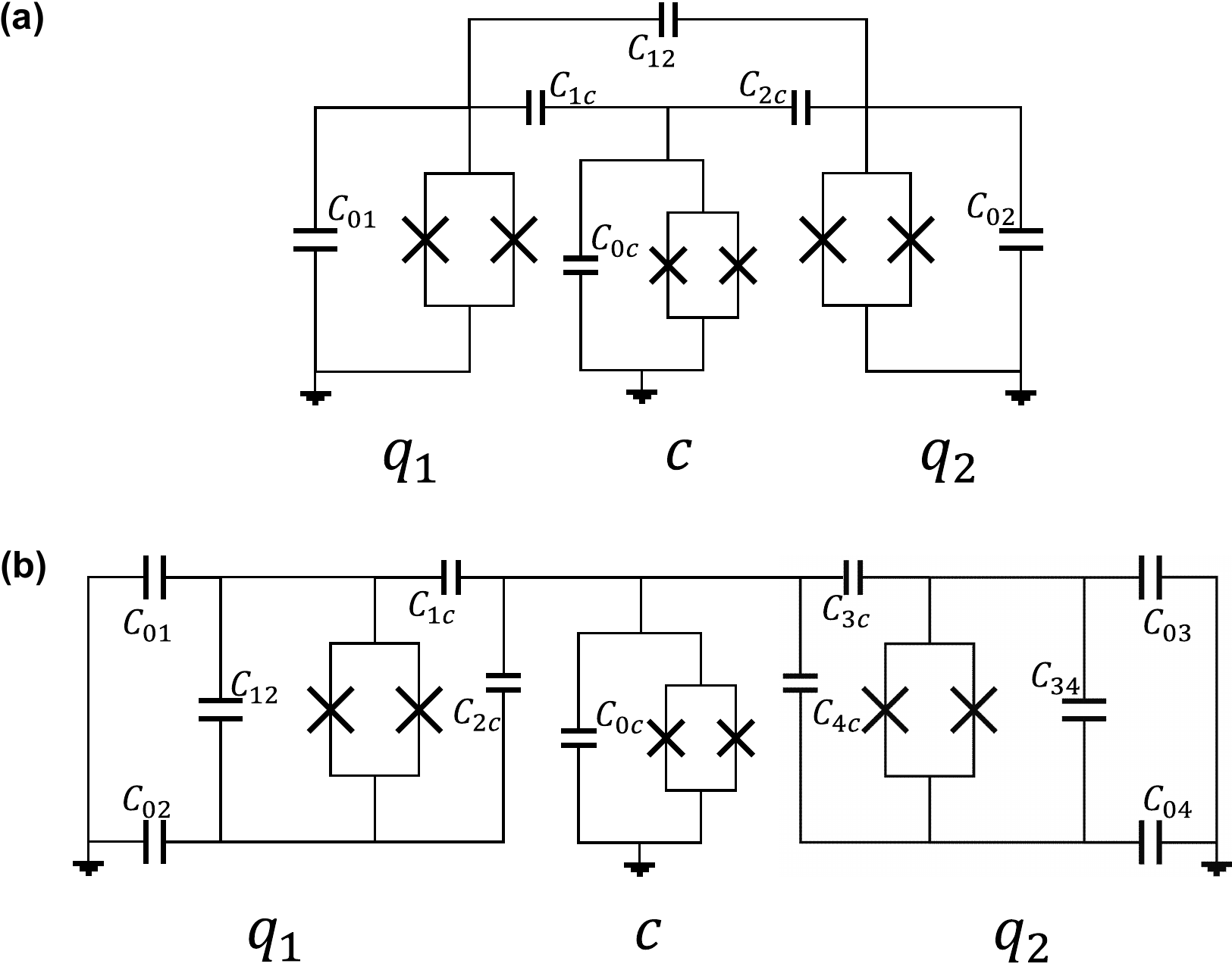}
    \caption{(a) The equivalent circuit of the G-G-G configuration. (b) The equivalent circuit of the F-G-F configuration, where we already omit the mutual capacitances between the two qubits $q_1$ and $q_2$.}
    \label{Scircuits}
\end{figure}

We now consider the general case in G-G-G configuration, as shown in Figure~\ref{Scircuits}(a). Then the capacitance matrix $\mathbf{C}$ reads
\begin{equation}\label{gggC1}
    \mathbf{C}=\mathbf{M}= 
    \begin{pmatrix}
    C_{01}+C_{1c}+C_{12} & -C_{1c} & -C_{12} \\
    -C_{1c} & C_{0c}+C_{1c}+C_{2c} & -C_{2c} \\
    -C_{12} & -C_{2c} & C_{02}+C_{2c}+C_{12} \\
    \end{pmatrix},
\end{equation}
with $\dot{\mathbf{\Phi}}=\dot{\mathbf{\Phi}}_M=\text{vec}(\dot{\Phi}_{\Delta_{q_1}},\dot{\Phi}_{\Delta_c},\dot{\Phi}_{\Delta_{q_2}})^T$. Then we obtain $E_{Cq_1}=e^2\mathbf{C}^{-1}_{11}/2, E_{Cc}=e^2\mathbf{C}^{-1}_{22}/2, E_{Cq_2}=e^2\mathbf{C}^{-1}_{33}/2, E_{q_1c}=e^2\mathbf{C}^{-1}_{12}, E_{q_2c}=e^2\mathbf{C}^{-1}_{23}, E_{q_1q_2}=e^2\mathbf{C}^{-1}_{13}$. Therefore, $A$ reads
\begin{equation}
    A_{\text{G-G-G}}=\frac{2E_{q_1q_2} E_{Cc}}{E_{q_1c}E_{q_2c}}=\frac{\mathbf{C}^{-1}_{13}\mathbf{C}^{-1}_{22}}{\mathbf{C}^{-1}_{12}\mathbf{C}^{-1}_{23}}.
\end{equation}

Taking the assumption of identical and symmetric qubits namely $C_{01}=C_{02}=C_{0q}, C_{1c}=C_{2c}=C_{qc}, C_{12}=C_{qq}$, then we can obtain the explicit expression of $A_{\text{G-G-G}}$ in the main text using \eqref{gggC1}
\begin{equation}\label{gggA}
    A_{\text{G-G-G}} = \frac{(C_{0q}+C_{qc})(C_{qc}^2+C_{0c}C_{qq}+2C_{qc}C_{qq})}{C_{qc}^2(C_{0q}+C_{qc}+2C_{qq})}.
\end{equation}
If we considering the long-range qubit-qubit coupling, namely taking $C_{qq}=0$, then one can immediately obtain $A_{\text{G-G-G}}=1$.

Also, we can further obtain 
\begin{align}
    B_{\text{G-G-G}} &=\frac{16E_{Cq}E_{Cc}}{E_{q_1c}E_{q_2c}}=\frac{4\mathbf{C}^{-1}_{11}\mathbf{C}^{-1}_{22}}{\mathbf{C}^{-1}_{12}\mathbf{C}^{-1}_{23}} \notag\\ 
    & =\frac{4 C_{0c}  (C_{0q}+C_{qc}) (C_{0q}+C_{qc}+C_{qq})}{C_{qc}^2 (C_{0q}+C_{qc}+2 C_{qq})} + \frac{4 (C_{0q}+C_{qc})  (2 C_{0q}+C_{qc}+2 C_{qq})}{C_{qc} (C_{0q}+C_{qc}+2 C_{qq})},\label{gggB} \\
    E_{Cq}/e^2 &=\mathbf{C}^{-1}_{11}/2=\frac{1}{4} \left(\frac{C_{0c}+2 C_{qc}}{C_{0c} (C_{0q}+C_{qc})+2 C_{0q} C_{qc}}  +\frac{1}{C_{0q}+C_{qc}+2 C_{qq}}\right).\label{Ecq}
\end{align}
With the help of \eqref{gggA}-\eqref{Ecq}, we can further realize the calculation in the main text to obtain the optimal set of capacitances in G-G-G configuration.

\subsection{F-G-F configuration}
In F-G-F configuration, the number of capacitances is enormous in the general case, for simplification we consider all the mutual capacitances between floating qubits as zero in advance, as shown in Figure~\ref{Scircuits}(b). Then its node potentials and Maxwell capacitance matrix read
\begin{equation}
\begin{aligned}
&\dot{\mathbf{\Phi}}_M=
\begin{pmatrix}
 \dot{\Phi}_{q_11} \\
 \dot{\Phi}_{q_12} \\
 \dot{\Phi}_{c} \\
 \dot{\Phi}_{q_21} \\
 \dot{\Phi}_{q_22}
\end{pmatrix},
\mathbf{M}=
\begin{pmatrix}
 \mathbf{M}_{11} & \mathbf{M}_{12} & \mathbf{M}_{13} & 0 & 0 \\
 \mathbf{M}_{21} & \mathbf{M}_{22} & \mathbf{M}_{23} & 0 & 0 \\
 \mathbf{M}_{31} & \mathbf{M}_{32} & \mathbf{M}_{33} & \mathbf{M}_{34} & \mathbf{M}_{35} \\
 0 & 0 & \mathbf{M}_{43} & \mathbf{M}_{44} & \mathbf{M}_{45} \\
 0 & 0 & \mathbf{M}_{53} & \mathbf{M}_{54} & \mathbf{M}_{55} \\
\end{pmatrix}, \\
\end{aligned}
\end{equation}
\begin{equation}
\begin{aligned}
&\mathbf{M}_{11}=C_{01}+C_{12}+C_{1c}, \\ 
&\mathbf{M}_{22}=C_{02}+C_{12}+C_{2c}, \\
&\mathbf{M}_{33}=C_{0c}+C_{1c}+C_{2c}+C_{3c}+C_{4c}, \\
&\mathbf{M}_{44}=C_{03}+C_{34}+C_{3c}, \\ 
&\mathbf{M}_{55}=C_{04}+C_{34}+C_{4c}, \\
&\mathbf{M}_{12}=\mathbf{M}_{21}=-C_{12}, \\
&\mathbf{M}_{13}=\mathbf{M}_{31}=-C_{1c}, \\
&\mathbf{M}_{23}=\mathbf{M}_{32}=-C_{2c}, \\
&\mathbf{M}_{34}=\mathbf{M}_{43}=-C_{3c}, \\
&\mathbf{M}_{35}=\mathbf{M}_{53}=-C_{4c}, \\
&\mathbf{M}_{45}=\mathbf{M}_{54}=-C_{34},
\end{aligned}
\end{equation}
where $\dot{\Phi}_{q_k1},\dot{\Phi}_{q_k2}, k=1,2$ represent the node potentials of two metal plates contained in the floating qubits, and $\dot{\Phi}_{c}$ is the node potential of the grounded coupler.
Using the transform matrix
\begin{equation}
    \mathbf{S}=\begin{pmatrix}
    1/2 & 1/2 & 0 & 0 & 0 \\
    1/2 & -1/2 & 0 & 0 & 0 \\
    0 & 0 & 1 & 0 & 0 \\
    0 & 0 & 0 & 1/2 & 1/2 \\
    0 & 0 & 0 & 1/2 & -1/2
    \end{pmatrix},
\end{equation}
we obtain 
\begin{equation}
\begin{aligned}
&\dot{\mathbf{\Phi}}=
\begin{pmatrix}
 \dot{\Phi}_{\Sigma_{q_1}} \\
 \dot{\Phi}_{\Delta_{q_1}} \\
 \dot{\Phi}_{\Delta_c} \\
 \dot{\Phi}_{\Sigma_{q_2}} \\
 \dot{\Phi}_{\Delta_{q_2}}
\end{pmatrix}, \quad
\mathbf{C}=
\begin{pmatrix}
 \mathbf{C}_{11} & \mathbf{C}_{12} & \mathbf{C}_{13} & 0 & 0 \\
 \mathbf{C}_{21} & \mathbf{C}_{22} & \mathbf{C}_{23} & 0 & 0 \\
 \mathbf{C}_{31} & \mathbf{C}_{32} & \mathbf{C}_{33} & \mathbf{C}_{34} & \mathbf{C}_{35} \\
 0 & 0 & \mathbf{C}_{43} & \mathbf{C}_{44} & \mathbf{C}_{45} \\
 0 & 0 & \mathbf{C}_{53} & \mathbf{C}_{54} & \mathbf{C}_{55} \\
\end{pmatrix}, \\
&\mathbf{C}_{11}= (C_{01}+C_{02}+C_{1c}+C_{2c})/4, \\ 
&\mathbf{C}_{22}= (C_{01}+C_{02}+4C_{12}+C_{1c}+C_{2c})/4, \\
&\mathbf{C}_{33}=C_{0c}+C_{1c}+C_{2c}+C_{3c}+C_{4c}, \\
&\mathbf{C}_{44}= (C_{03}+C_{04}+C_{3c}+C_{4c})/4, \\ 
&\mathbf{C}_{55}= (C_{03}+C_{04}+4C_{34}+C_{3c}+C_{4c})/4, \\
&\mathbf{C}_{12}=\mathbf{C}_{21}= (C_{01}-C_{02}+C_{1c}-C_{2c})/4, \\
&\mathbf{C}_{13}=\mathbf{C}_{31}= (-C_{1c}-C_{2c})/2, \\
&\mathbf{C}_{23}=\mathbf{C}_{32}=(C_{2c}-C_{1c})/2, \\
&\mathbf{C}_{34}=\mathbf{C}_{43}= (-C_{3c}-C_{4c})/2, \\
&\mathbf{C}_{35}=\mathbf{C}_{53}=(C_{4c}-C_{3c})/2, \\
&\mathbf{C}_{45}=\mathbf{C}_{54}= (C_{03}-C_{04}+C_{3c}-C_{4c})/4.
\end{aligned}
\end{equation}
Therefore, the charging and coupling energies are $E_{Cq_1}=e^2\mathbf{C}^{-1}_{22}/2, E_{Cc}=e^2\mathbf{C}^{-1}_{33}/2, E_{Cq_2}=e^2\mathbf{C}^{-1}_{55}/2, E_{q_1c}=e^2\mathbf{C}^{-1}_{23}, E_{q_2c}=e^2\mathbf{C}^{-1}_{35}, E_{q_1q_2}=e^2\mathbf{C}^{-1}_{25}$, resulting in
\begin{equation}
    A_{\text{F-G-F}}=\frac{\mathbf{C}^{-1}_{25} \mathbf{C}^{-1}_{33}}{\mathbf{C}^{-1}_{23}\mathbf{C}^{-1}_{35}}.
\end{equation}
Through mathematical calculation with aid of software, $ A$ is evaluated as one as well in F-G-F configuration.

\section{Derivation of turned-on qubit-qubit effective coupling strength and related parameters}
\label{section4}

We start with the QQe coupling 
\begin{equation} \label{g1}
    g=\frac{2\omega_q}{B}\left(A-\frac{\omega_c^2}{\omega_c^2-\omega_q^2}\right).
\end{equation}
In some cases, we care about the qubit-coupler dispersive rate $\beta$ rather than the coupler frequency $\omega_c$, therefore, by the definition \eqref{beta} we obtain
\begin{equation}
    \frac{\omega_q}{\omega_c} = 1-\frac{\beta\left(\sqrt{\beta^2+4\vert B\vert }-\beta\right)}{2\vert B\vert },
\end{equation}
and then \eqref{g1} is rewritten as
\begin{equation}\label{g2}
    g = \frac{2\omega_q}{B}\left[A-\frac{1}{1-\left(1-\frac{\beta\left(\sqrt{\beta^2+4\vert B\vert }-\beta\right)}{2\vert B\vert }\right)^2}\right].
\end{equation}

Using (\ref{g1}) and (\ref{g2}), we can tie up the critical coupler frequencies $\omega_{\text{on}}, \omega_{\text{off}}$ and consequently count in three constraints (namely the zero-coupling requirement, the qubit-coupler dispersive coupling, and the limited coupler frequency) altogether on the way studying the turned-on QQe coupling.

For  the sub-regime $\omega_{\text{on}}<\omega_{\text{off}}$, we have $\omega_s = \omega_{\text{on}}$ and $\omega_l = \omega_{\text{off}}$. We first use (\ref{g1}) at $\omega_l= \omega_{\text{off}}$ and let $g=0$, leading to 
\begin{equation}\label{A1}
    A = \frac{\omega_l^2}{\omega_l^2-\omega_q^2}.
\end{equation}
While at $\omega_s = \omega_{\text{on}}$ where $\beta_s$ matters more than $\omega_s$, we substitute (\ref{A1}) into  (\ref{g2}), leading to 
\begin{equation}\label{Sgon}
    \left\vert g_{\text{on}}\right\vert =\frac{2\omega_q}{\vert B\vert }\left[\frac{1}{1-\left(1-\frac{\beta_s\left(\sqrt{\beta_s^2+4\vert B\vert }-\beta_s\right)}{2\vert B\vert }\right)^2}-\frac{\omega_l^2}{\omega_l^2-\omega_q^2}\right].
\end{equation}

For the regime $\omega_{\text{on}}>\omega_{\text{off}}$, we use (\ref{g2}) at $\omega_s= \omega_{\text{off}}$, leading to 
\begin{equation}\label{A2}
    A = \frac{1}{1-\left(1-\frac{\beta_s\left(\sqrt{\beta_s^2+4\vert B\vert }-\beta_s\right)}{2\vert B\vert }\right)^2},
\end{equation}
then substituting (\ref{A2}) into (\ref{g1}) and taking $\omega_{c}=\omega_l$, we obtain (\ref{Sgon}) as well.
As a result, the two regimes can be considered uniformly while distinguished by different $A$.

To simplify \eqref{Sgon}, we further introduce 
\begin{equation}\label{xy}
    \begin{aligned}
       x  = \frac{\omega_q}{\omega_s} =&  1-\frac{\beta_s\left(\sqrt{\beta_s^2+4\vert B\vert }-\beta_s\right)}{2\vert B\vert }, \\ \; \\
       y  = \frac{\omega_s}{\omega_l} =&  \frac{1}{x}\frac{\omega_q}{\omega_l},
    \end{aligned}
\end{equation}
where $x,y$ are both dimensionless. Since we always have $\omega_q<\omega_s<\omega_l$ in $\omega_q<\omega_c$ regime, $x,y$ both lie in $(0,1)$.
Therefore, the turned-on QQe coupling strength (\ref{Sgon}) can be finally derived as
\begin{equation}
    \left\vert g_{\text{on}}\right\vert  = \frac{2 \omega _l}{\beta _s^2}\left[\frac{(1-x)\left(1-y^2\right) x^2 y }{(1+x) \left(1-x^2 y^2\right)}\right],
\end{equation}
where the restraints $\omega_l$ and  $\beta_s$  are both extracted as detached pre-factors, leading to a concise presentation.

In addition, the related quantities $A,B,\omega_q, \omega_s$ can also be expressed in terms of  $x, y, \omega_l, \beta_s$. In particular, using (\ref{xy}) we obtain
\begin{eqnarray}
\vert B\vert  &&= \frac{x}{(1-x)^2}\beta_s^2, \label{B} \\
\omega_q &&= xy\omega_l, \label{wq} \\
\omega_s &&= y\omega_l.
\end{eqnarray}
Significantly, $B,\omega_q,\omega_s$ are the same for both regimes. Using \eqref{A1},~\eqref{A2},~\eqref{xy}, we further obtain $A$ for the two different sub-regimes, expressing as
\begin{equation}
A=\left\{
    \begin{array}{cc}
        \frac{1}{1-x^2}, &  \quad \omega_{\text{on}} > \omega_{\text{off}},\\
        \frac{1}{1-x^2y^2}, &  \quad \omega_{\text{on}} < \omega_{\text{off}}.
    \end{array}
    \right. 
\end{equation}

\section{The sign  of $E_{12}$}
\label{section5}
In the main text, we stressed that $B$ shares the same sign with $E_{12}$ (the coupling energy between two qubits). Next, let us look into the sign of $E_{12}$.

Taking G-G-G configuration as an example, by \eqref{gggC1} we can obtain 
\begin{equation}\label{E12p}
    E_{12}/e^2=\mathbf{C}_{13}^{-1}= 
    \frac{C_{0c} C_{qq}+C_{qc}^2+2 C_{qc} C_{qq}}{(C_{0c} (C_{0q}+C_{qc})+2 C_{0q} C_{qc}) (C_{0q}+C_{qc}+2 C_{qq})},
\end{equation}
where we take the assumption of identical and symmetric qubits. Apparently, $E_{12}$ is always positive. However, if we change the sign of an arbitrary node flux by definition, e.g., redefining $\Phi_{q_1}' \equiv -\Phi_{q_1}$, then the sign of corresponding mutual capacitances in Maxwell matrix is reversed, leading to 
\begin{equation}\label{gggC2}
    \mathbf{C}=\mathbf{M}= 
    \begin{pmatrix}
    C_{01}+C_{1c}+C_{12} & C_{1c} & C_{12} \\
    C_{1c} & C_{0c}+C_{1c}+C_{2c} & -C_{2c} \\
    C_{12} & -C_{2c} & C_{02}+C_{2c}+C_{12} \\
    \end{pmatrix}.
\end{equation}
In this case, we obtain 
\begin{equation}
    E_{12}/e^2=\mathbf{C}_{13}^{-1}= 
    -\frac{C_{0c} C_{qq}+C_{qc}^2+2 C_{qc} C_{qq}}{(C_{0c} (C_{0q}+C_{qc})+2 C_{0q} C_{qc}) (C_{0q}+C_{qc}+2 C_{qq})},
\end{equation}
which is exactly the negative value of \eqref{E12p}. This implies the sign of $E_{12}$ can be always managed artificially by arbitrarily adjusting the sign of node fluxes by definition. 

For some complicated QCQ systems, However, the sign of $E_{12}$ is hard to tell or we are not even able to obtain a concise presentation of $E_{12}$ to distinguish its sign, much less to manage it. Therefore, in these cases,  we need to take both cases of $E_{12}$ to decide $B$ for further discussion.

\section*{Reference}
\bibliography{main}

\end{document}